# Etude des effets non linéaires observés sur les oscillations d'un pendule simple

Par Thomas Gibaud [1] et Alain Gibaud [2]

1- Université de Fribourg, Département de Physique, Chemin du Musée 3, 1700 Fribourg, Suisse. thomas.gibaud@unifr.ch

2- Université du Maine, Laboratoire de Physique de l'Etat Condensé, Faculté des Sciences, UMR 6087 CNRS, 72085 Le Mans Cedex 09.

## *RÉSUMÉ*

*Nous présentons dans cet article une étude des effets non linéaires engendrés par l'anharmonicité du potentiel du pendule simple. Dans un rappel théorique nous mettons en évidence que l'anharmonicité du potentiel engendre des harmoniques supplémentaires et le non isochronisme des oscillations. Ces phénomènes sont d'autant plus importants que l'on s'écarte des oscillations aux petits angles, le régime de validité de l'approximation harmonique. La mesure est appréhendée au moyen du boîtier d'acquisition SYSAM-SP5 couplé au logiciel Latis pro et du pendule commercialisé par Eurosmart. Nous montrons que seule une analyse fine par simulation de la courbe enregistrée permet d'obtenir une précision suffisante pour décrire l'évolution quadratique de la période en fonction de l'amplitude des oscillations. Nous constatons que nous pouvons détecter les harmoniques supplémentaires dans les oscillations lorsque l'amplitude devient très élevée.*

## 1. Introduction

Le pendule simple appartient à cette famille de systèmes physiques qui déplacés légèrement de leur position d'équilibre se mettent à osciller autour de cette position d'équilibre. De tels systèmes sont très communs en physique aussi bien en mécanique (pendule, ressort… [1]) qu'en électricité (circuit *RLC* série) ou encore en physique du solide (vibration d'un atome dans le réseau cristallin [2,3]). Pour peu que le mouvement d'oscillation soit de faible amplitude, le système se comporte comme un oscillateur linéaire encore appelé oscillateur harmonique. Mais dès que l'amplitude des oscillations devient grande alors le système devient un oscillateur non linéaire.

L'objectif de cet article est d'aborder d'abord théoriquement puis expérimentalement la non linéarité en prenant l'exemple du pendule simple. Dans un premier temps nous montrons que les effets non linéaires, à savoir la présence d'harmoniques supplémentaires dans le régime des oscillations et la dépendance de la période en fonction de l'amplitude maximale des oscillations, découlent de l'anharmonicité du potentiel du pendule. Dans la 2ième et 3ième partie nous nous attachons à mettre en évidences de façon expérimentale ces deux effets non linéaires. Comme ces effets correspondent à de faibles corrections du modèle harmonique, nous attachons une importance capitale à la mesure et à la précision avec laquelle elle est effectuée.

## 2. Théorie : origine et conséquences des effets non linéaires

Le système étudié est un pendule simple de longueur, $l$, de masse, $m$. L'amplitude des oscillations est paramétrée par l'angle, $\theta$, entre la verticale et la tige du pendule. L'angle est mesuré grâce un potentiomètre qui transforme le signal mécanique en signal électrique. La masse est suffisamment éloignée de l'axe de rotation pour traiter le problème dans l'approximation de la mécanique du point, ce qui revient à admettre que le pendule est un pendule simple.

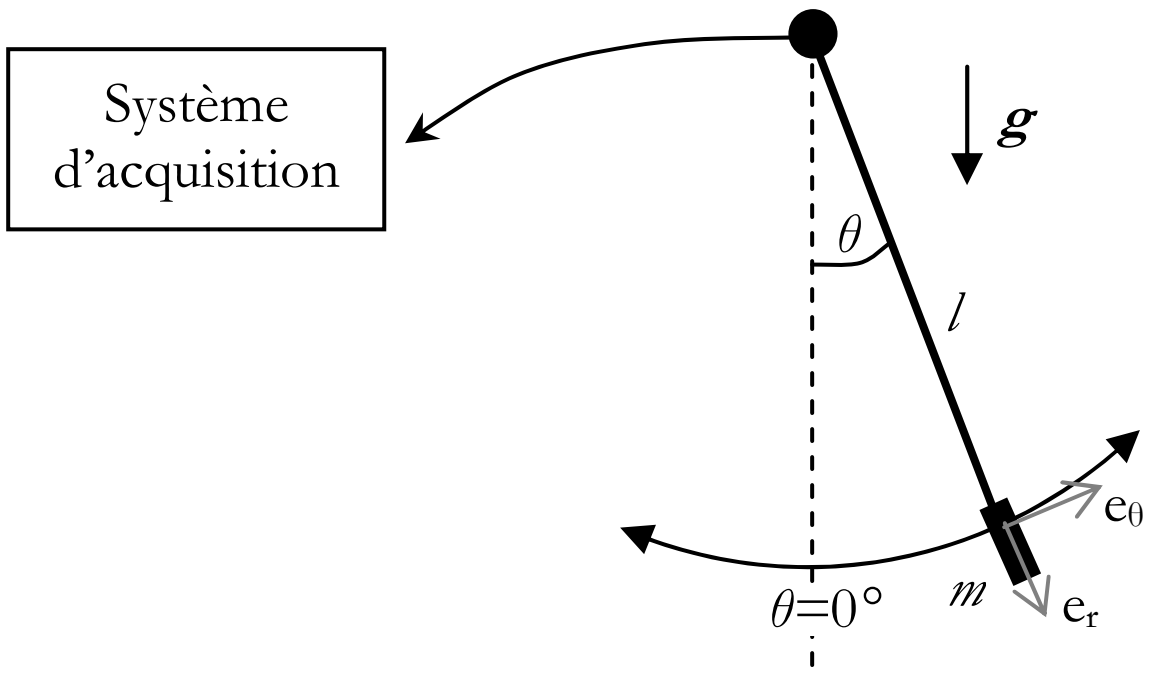

Figure 1 : Schéma du pendule. $\theta$=0° correspond à la position d'équilibre stable du pendule.

L'étude des effets non linéaires jouant sur les oscillations d'un pendule simple s'inscrit dans la thématique plus générale des effets de non linéarité dus à un potentiel anharmonique [4,5]. Un potentiel est harmonique s'il est quadratique en fonction du paramètre décrivant les variations de position du système autour de sa position d'équilibre. Il est anharmonique dans le cas contraire. L'exemple typique du potentiel harmonique est celui du pendule élastique pour lequel $E_p = \frac{1}{2}kx^2$, x étant le degré de liberté du système. Un potentiel peut souvent être approximé par un potentiel harmonique pourvu que le système ne s'écarte pas trop de sa position d'équilibre stable. Lorsque l'amplitude des oscillations du pendule simple devient trop importante l'approximation de l'énergie potentielle par un potentiel harmonique est insuffisante. Examinons le cas du pendule simple.

Un pendule simple en absence de frottement est mécaniquement isolé. L'énergie mécanique se conserve. Comme le système ne possède qu'un degré de liberté, la dérivée de l'énergie mécanique par rapport au temps conduit directement à l'équation différentielle du mouvement :

$$\begin{cases} \text{Energie cinétique : } E_c\left(\dot{\theta}\right) = \frac{1}{2} m \left(l\dot{\theta}\right)^2 \\ \text{Energie potentielle : } E_p\left(\theta\right) = mgl\left(1 - \cos\theta\right) \\ \text{Convention: } E_p(\theta = 0°) = 0 \end{cases}$$

$$\Rightarrow E = E_c + E_p = cste \Leftrightarrow \ddot{\theta} + \omega_0^2 \sin\theta = 0, \text{ avec } \omega_0 = \sqrt{\frac{g}{l}}$$

L'équation du mouvement est non linéaire du fait de la présence du sinus qui dérive de l'énergie potentielle. Pour résoudre une telle équation, de façon générale, on développe l'énergie potentielle en série de Taylor autour de la position d'équilibre. Plus on prend en considération de termes de puissance élevée, meilleure est l'adéquation entre le potentiel et le développement. On peut alors espérer décrire le comportement du pendule pour des amplitudes d'oscillations élevées.

$$E_p(\theta) \approx \underbrace{E_p(0)}_{\substack{0 \text{ par} \\ \text{convention}}} + \underbrace{\left.\frac{dE_p}{d\theta}\right)_{\theta=0}}_{\substack{0 \text{ car le} \\ \text{potentiel} \\ \text{présente} \\ \text{un minimum} \\ \text{en } \theta=0}} \theta + \underbrace{\frac{1}{2}\left.\frac{d^2E_p}{d\theta^2}\right)_{\theta=0}}_{\substack{\text{approximation} \\ \text{harmonique} \\ \text{du potentiel}}} \theta^2 + \underbrace{\frac{1}{6}\left.\frac{d^3E_p}{d\theta^3}\right)_{\theta=0}}_{\substack{0 \text{ car le potentiel} \\ \text{est pair} \\ E_P(\theta)=E_P(-\theta)}} \theta^3 + \underbrace{\frac{1}{24}\left.\frac{d^4E_p}{d\theta^4}\right)_{\theta=0}}_{1^{er} \text{ terme non linéaire}} \theta^4 + \ldots$$

Dans l'approximation harmonique où approximation des petits angles, on s'arrête dans le développement de l'énergie potentielle au terme d'ordre 2. L'équation différentielle précédente s'identifie alors à celle d'un oscillateur harmonique :

$$\theta \text{ petit} \Rightarrow \begin{cases} \text{potentiel harmonique : } E_p = mgl\dfrac{\theta^2}{2} + o\left(\theta^2\right) \\ \text{force de rappel : } \mathbf{F} = -\dfrac{1}{l}\dfrac{dE_p}{d\theta}\mathbf{e}_\theta = -mg\theta\mathbf{e}_\theta \\ \text{équation linéaire : } \ddot{\theta} + \omega_0^2\theta = 0 \end{cases}$$

La période est alors indépendante de l'amplitude des oscillations et s'écrit :

$$T_0 = \frac{2\pi}{\omega_0} = 2\pi\sqrt{\frac{l}{g}} = cste$$

Si θ devient grand l'approximation précédente n'est plus pertinente. Il faut pousser le développement plus loin pour obtenir une meilleure description de l'énergie potentielle. On se limite à l'ordre 4, l'idée étant que, plus on effectue un développement en puissance de θ élevée, plus on se rapproche du potentiel réel.

$$\theta \text{ grand} \Rightarrow \begin{cases} \text{potentiel anharmonique : } E_p = mgl\left[\dfrac{\theta^2}{2} - \dfrac{\theta^4}{24} + o\left(\theta^4\right)\right] \\ \text{force de rappel : } \mathbf{F} = -\dfrac{1}{l}\dfrac{dE_p}{d\theta}\mathbf{e}_\theta = -mg\left[\theta - \dfrac{\theta^3}{6}\right]\mathbf{e}_\theta \\ \text{équation non linéaire : } \ddot{\theta} + \omega_0^2\left(\theta - \dfrac{\theta^3}{6}\right) = 0 \end{cases}$$

L'équation différentielle est connue sous le nom d'équation de Duffing sans force motrice et peut être résolue en utilisant la méthode perturbative de Poincaré-Lindstedt [6-8]. Contrairement au cas de l'oscillateur harmonique la force n'est plus une fonction linéaire de la variable θ. C'est d'ailleurs l'origine de l'appellation de cette classe de phénomènes : phénomène non linéaire. Cette non linéarité n'est pas sans conséquences. Le principe de superposition des solutions n'est désormais plus valable. Il en résulte qu'on ne peut plus appliquer la méthode traditionnelle des nombres complexes pour résoudre l'équation différentielle.

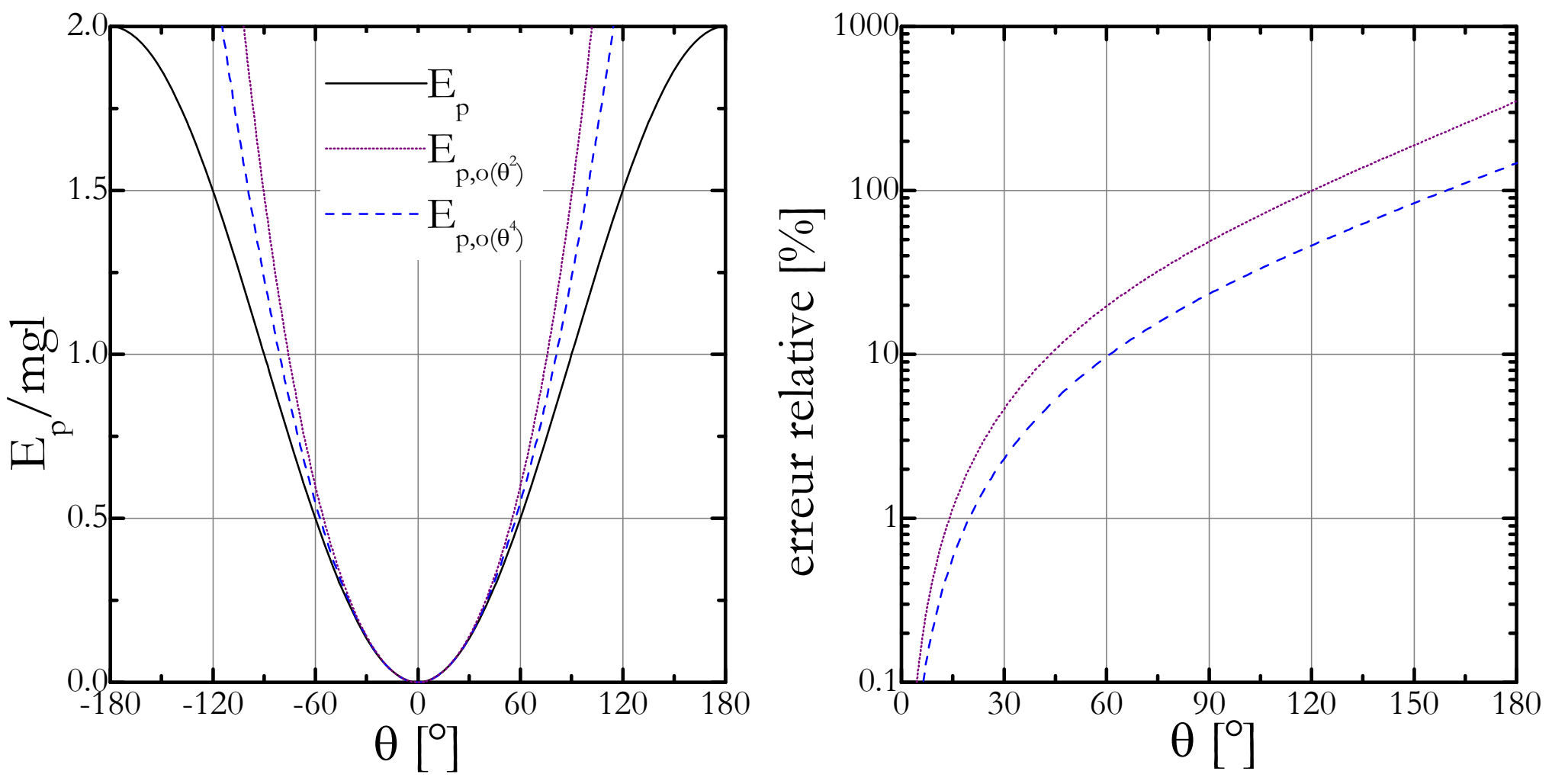

Figure 2 : (a) comparaison entre l'expression exacte de l'énergie potentielle et l'approximation harmonique, développement au 2nd ordre et le développement au 4ième ordre. (b) Erreur relative entre l'expression exacte de l'énergie potentielle et l'approximation harmonique et le développement au 4ième ordre.

Si l'on remplace θ(t) par une solution sinusoïdale de pulsation ω on constate que le terme en $\theta^3$ dans l'équation engendre des harmoniques d'ordre 3, [2,5,9].

$$\sin^3 \omega t = \frac{3\sin\omega t - \sin 3\omega t}{4}$$

D'où l'idée de chercher une solution du type θ(t)= $\theta_0$.[sinωt+ε.sin3ωt]. Cette méthode est dite *perturbative.* La solution est la somme d'un terme sinusoïdal et d'un terme correctif pour lequel ε <<1. Il est effectivement plus simple mathématiquement de procéder par corrections successives que de s'attaquer directement à la solution sans approximation souvent incalculable de façon littérale. La résolution de l'équation différentielle non linéaire se fait en y injectant la solution préconisée.

$$\begin{cases} \text{Equation : } \ddot{\theta} + \omega_0^2\left(\theta - \dfrac{\theta^3}{6}\right) = 0 \\ \text{Solution préconisée : } \theta(t) = \theta_0\left[\sin\omega t + \varepsilon\sin 3\omega t\right] \end{cases}$$

$$\Rightarrow -\omega^2\theta_0\left[\sin\omega t + 3\varepsilon\sin 3\omega t\right] + \omega^2\theta_0\left[\sin\omega t + 3\varepsilon\sin 3\omega t\right] - \frac{\omega^2\theta_0^3}{6}\sin^3\omega t = 0 + o(\varepsilon)$$

En regroupant les termes de même harmonicité et en identifiant les préfacteurs, on trouve :

$$\sin\omega t\left[\omega_0^2 - \omega^2\right] + \sin 3\omega t\left[\varepsilon\left(\omega_0^2 - 9\omega^2\right) - \frac{\omega^2\theta_0^3}{24}\right] = 0 + o(\varepsilon)$$

$$\Rightarrow \begin{cases} T = T_0\left(1 + \dfrac{\theta_0^2}{16} + o\left(\theta_0^2\right)\right) \text{ avec } T_0 = 2\pi\sqrt{\dfrac{l}{g}} \\ \varepsilon = \dfrac{\theta_0^2}{192} \end{cases}$$

Les effets non linéaires sur le pendule simple se manifestent doublement. L'anharmonicité du potentiel engendre des harmoniques supplémentaires et le non isochronisme des oscillations. Nous rappelons à ce titre que les oscillations sont dites isochrones si leur période est indépendante de l'amplitude. On voit très bien que les oscillations ne peuvent être isochrones que pour des amplitudes très faibles.

Il s'agit désormais de mettre en évidence expérimentalement les effets non linéaires observables sur le pendule, à savoir, la variation de la période avec l'amplitude maximale des oscillations et la présence d'harmoniques. Comme ces phénomènes correspondent à de faibles corrections par rapport à ce que l'on peut attendre du modèle de l'oscillateur harmonique, il est pertinent de se demander si on peut détecter expérimentalement ces corrections. La précision et l'incertitude avec laquelle les mesures sont effectuées sont donc centrales dans ce travail expérimental. Elles dépendent du matériel utilisé : le boîtier d'acquisition SYSAM-SP5, le logiciel Latis pro et le pendule commercialisé par Eurosmart.

## 3. Expérience : étude du non isochronisme des oscillations

D'après la partie théorique, la variation absolue de période du pendule évolue de façon quadratique par rapport à l'amplitude maximale des oscillations, $\theta_0$ :

$$T - T_0 = T_0\frac{\theta_0^2}{16}$$

Dans cette expression, l'amplitude $\theta_0$ s'exprime en radians ce qui montre que la variation absolue de période reste faible (on rappelle que $\theta_0(rad) \approx \theta_0(deg)/60$). Il est donc utile de prendre un pendule le plus long possible pour que les variations de période en fonction de l'amplitude soient plus facilement détectables.

La question importante qui se pose ensuite concerne la précision expérimentale avec laquelle la période peut être mesurée. A titre d'exemple, la variation relative de la période n'est approximativement que de 2% lorsque $\theta_0$=30°. Il faut donc pouvoir mesurer la période à mieux que le 0.1% si l'on veut espérer mesurer l'effet de non isochronisme. A ceci s'ajoute le problème de l'amortissement du pendule. L'amortissement par frottement visqueux induit une augmentation de la période d'oscillation du pendule [10]. Pour un pendule donné (et à faible vitesse) cet effet est constant ce qui permet de l'intégrer à $T_0$.

La Figure 3a montre la variation de l'amplitude en fonction du temps. Une vue agrandie correspondant à plusieurs pseudo-périodes est présentée sur la Figure 3b et une vue détaillée d'une oscillation est donnée Figure 3c. De cette dernière figure, on peut constater que l'incertitude sur la mesure est double et intimement lié à la numérisation du signal. Elle provient à la fois du pas temporel d'acquisition, encore appelé période d'échantillonnage, *Te* et du quantum, $q$=0.35°qui est la plus petite variation angulaire mesurable. Dans cet article toutes les acquisitions ont été réalisées avec une période d'échantillonnage *Te*=1ms.

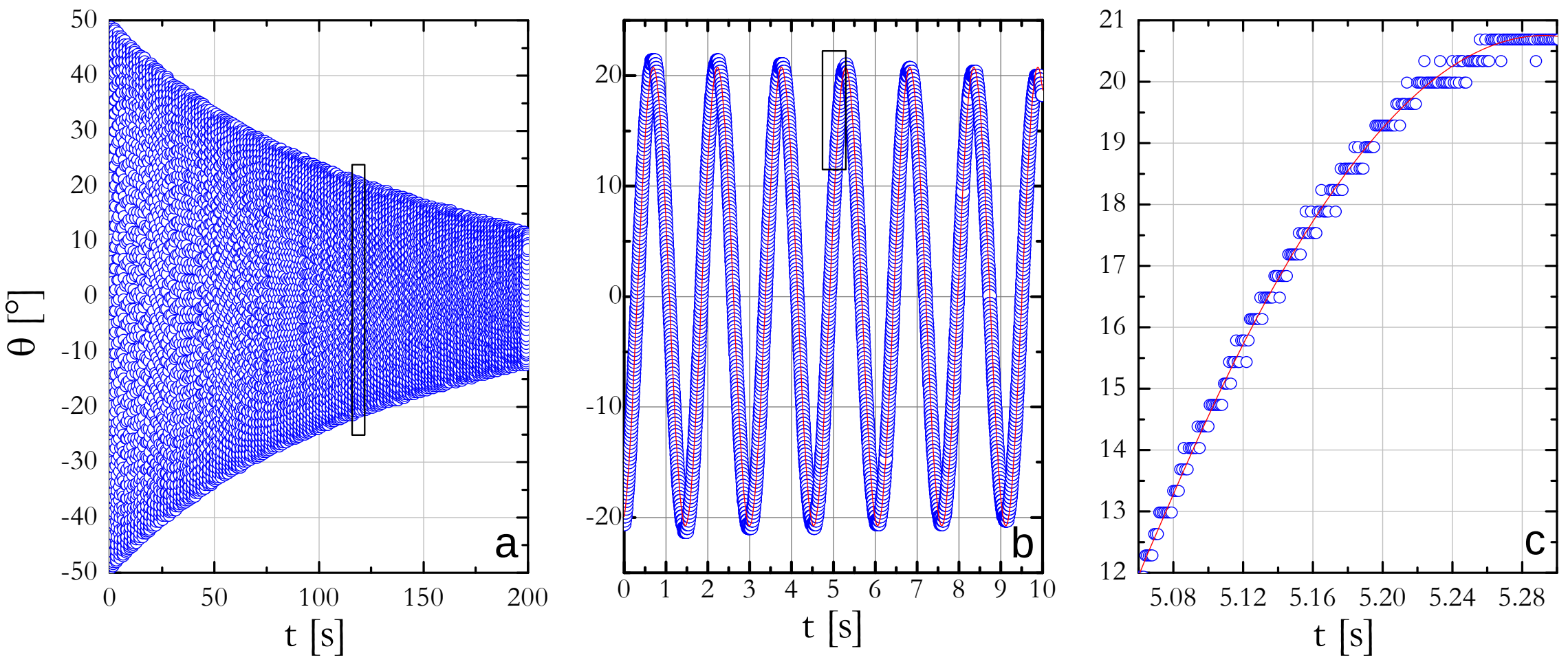

Figure 3 : (a) variation de l'amplitude en fonction du temps. (b) zoom sur l'amplitude $\theta_0$=20.8°. La courbe en trait plein représente l'ajustement d'une sinusoïdale sur les points expérimentaux. (c) zoom sur les détails de la courbe permettant de mettre en évidence les paramètres d'acquisition : quantum et échantillonnage.

A partir de cette expérience, on peut dégager trois méthodes permettant d'obtenir la période. La première consiste à effectuer la transformée de Fourier du signal. Cette méthode est toutefois inadaptée car comme on le voit les oscillations sont amorties ce qui impose de limiter le temps de mesure à quelques périodes pour avoir une amplitude constante. Cette limitation sur le temps total d'acquisition est un frein à la précision de la mesure de la fréquence d'oscillation puisque l'incertitude sur la fréquence est égale à l'inverse du temps d'acquisition Le pas fréquentiel dans l'espace de Fourier est donné par *1/NTe*, ou *Te* est la période d'échantillonnage et *N* le nombre de points de mesure. Il s'ensuit que le pas fréquentiel est donné par l'inverse du plus grand temps mesuré lors de l'acquisition, à savoir la durée totale d'acquisition *NTe*. A titre d'exemple, pour un temps d'acquisition de 10s l'amplitude est quasi constante mais la précision sur la fréquence des oscillation dans l'espace de Fourier n'est que de 0.1Hz soit environ 15% d'erreur relative : $f$(20.8°)=0.7±0.1 Hz. La seconde méthode consiste à mesurer 5 périodes avec le curseur. On se place non sur la crête de l'oscillation mais sur le milieu de l'oscillation où la pente est la plus forte pour mesurer la période le plus précisément possible. Cette méthode est presque convenable puisque sur l'exemple choisi, on obtient la période à 0.2% près : $T$(20.8±0.2°)=1.532±0.003s. L'incertitude est plus grande que la période d'échantillonnage, *Te*. En effet, comme on le voit sur la figure 3c, la fonction varie par palier de hauteur, le quantum, car le temps nécessaire à l'amplitude pour varier d'un quantum est plus long que *Te*. La dernière méthode consiste à ajuster un modèle de sinusoïdal pour déterminer la période. Lors de l'ajustement, voir graphe 3b, on prend en compte les incertitudes liées à l'acquisition pour évaluer les incertitudes sur les paramètres ajustés, à savoir l'amplitude, la phase à l'origine et la période. Cette méthode est convenable puisque sur l'exemple choisi on obtient la période à 0.04% près : $T$(20.8±0.2°)=1.5314±0.0005s.

Nous avons testé les 2 dernières méthodes de mesure de la période : la méthode du « curseur » et la méthode de « l'ajustement sinusoïdal ». Les résultats figure sur le graphe 4. La méthode du curseur présente une ambiguïté. Effectivement, les incertitudes sont telles qu'aussi bien un ajustement linéaire que l'ajustement du modèle non linéaire s'adapte aux mesures. Comme le système oscille autour de sa position d'équilibre : $dE_p/d\theta=0$. Un ajustement linéaire n'est donc pas de mise, mais d'un point de vue purement expérimental, les mesures ne sont pas assez précises pour infirmer ou confirmer le modèle. Par contre, on constate que la dernière méthode est suffisamment précise pour valider sans ambiguïté l'effet non linéaire et son modèle. Les résultats sont en effet en parfaite adéquation avec le modèle pour des angles jusqu'à 30°. Une méthode quantitative pour évaluer la pertinence de l'ajustement consiste à regarder le $\chi^2$. On rappel que le $\chi^2$ est la grandeur que l'on minimise dans un ajustement de « type moindre carré ». Un ajustement est réussi si le $\chi^2$ converge vers le $\chi^2$ théorique donné par $\chi^2_{\text{Théorique}}$=n-k ou n est le nombre de points expérimentaux et k est le nombre de paramètres de l'ajustement (ici, n=11 et k=2). Le $\chi^2$ est un peu faible par rapport au $\chi^2$ théorique, 9, signe que l'on a probablement un peu surévalué les incertitudes.

$$\text{Modèle : } T = T_0\left(1 + a\frac{\theta_0^2}{16}\right)$$

$$\text{Ajustement : }\left|\begin{array}{l} T_0 = 1.5191 \pm 0.0003 s \\ a = 1.03 \pm 0.03 \\ \chi^2 = 5.8 \end{array}\right.$$

Au-delà de 30°, la période mesurée est supérieure à celle prédite par le modèle. Cet écart varie dans le bon sens. Effectivement dans le modèle, nous nous sommes limité au second ordre et les ordres supérieurs on une contribution additive sur la période, [2, 5].

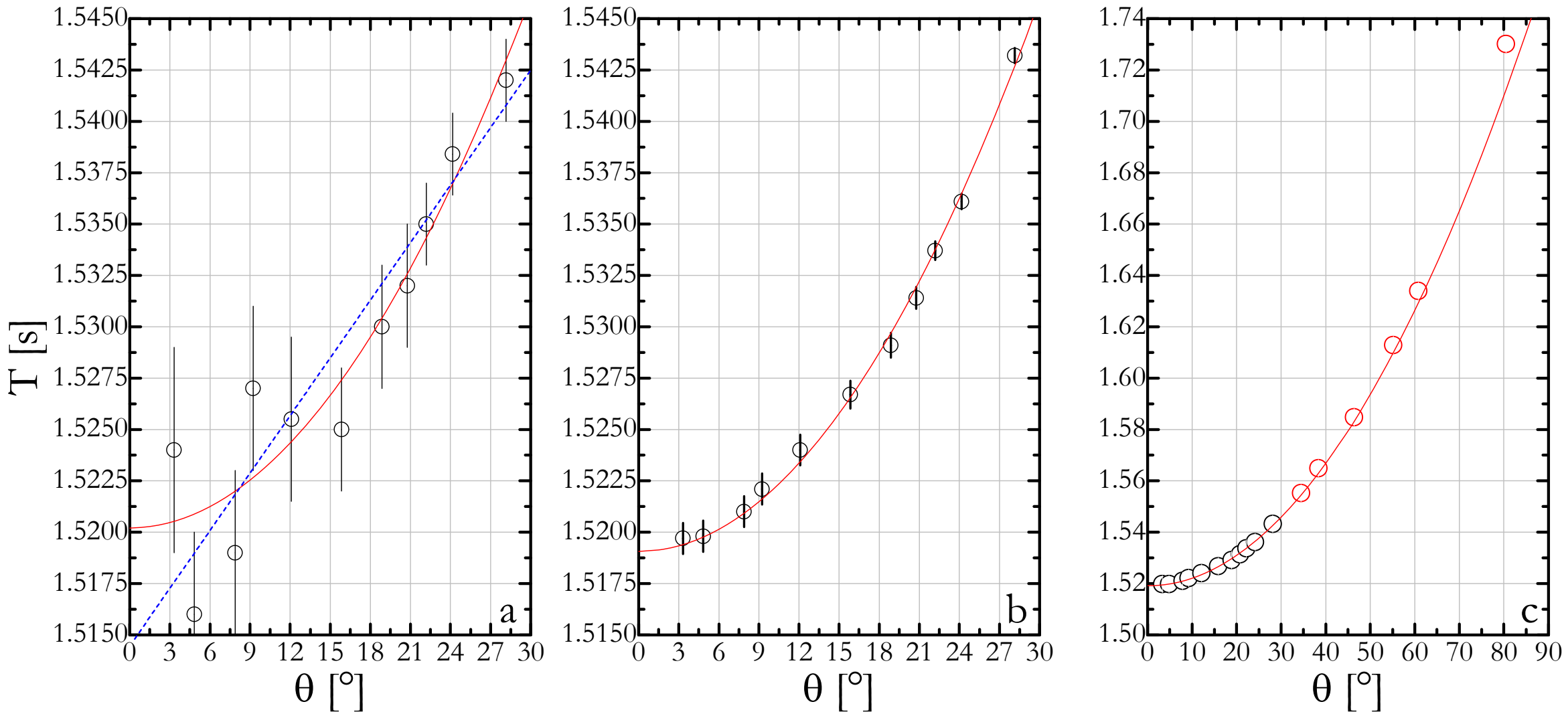

Figure 4 : (a) Mesure de la période selon la méthode du « curseur ». La courbe en trait plein représente l'ajustement du modèle non linéaire. La courbe en pointillé représente un ajustement linéaire (b) Mesure de la période selon la méthode de « l'ajustement sinusoïdal ». La courbe en trait plein représente l'ajustement du modèle non linéaire. Les incertitudes sont plus importantes aux faibles amplitudes car le rapport entre le quantum et l'amplitude est plus faible donc l'ajustement de la sinusoïde ou le pointage avec le curseur est moins précis. (c) Variation de la période en fonction de l'amplitude sur une plage angulaire plus étendue. La courbe en trait plein représente toujours le modèle.

## 4. Expérience : mise en évidence de l'harmonique d'ordre 3

La détection de l'harmonique d'ordre 3 est encore plus délicate car le préfacteur du terme correctif, est en $\varepsilon=(\theta_0)^2/192$ c'est-à-dire plus de 10 fois plus petit que la correction équivalente sur la période. Compte tenu du quantum imposé par la carte d'acquisition, la mesure est effectuée à 0.35° près, il faut donc aller à des angles bien plus grands que $\theta_0$=60°, angle à partir duquel $\varepsilon$=0.35°. On a aussi intérêt à tracer la vitesse angulaire car l'amplitude des harmoniques y est supérieure d'un facteur trois comme on le voit sur les expressions théoriques au 1$^{\text{er}}$ ordre en $\varepsilon$ :

$$\begin{cases} \theta(\mathrm{t}) = \theta_0[\sin\omega\mathrm{t} + \varepsilon\sin 3\omega\mathrm{t}] + \mathrm{o}(\varepsilon) \\ \dot{\theta}(\mathrm{t}) = \omega\theta_0[\cos\omega\mathrm{t} + 3\varepsilon\cos 3\omega\mathrm{t}] + \mathrm{o}(\varepsilon) \end{cases}$$

On obtient la vitesse angulaire en dérivant l'amplitude par rapport au temps. Comme le signal est numérique, il y a des discontinuités locales d'amplitude (voir figure 3c), le quantum, qui engendrent des pics dans la dérivée. Ces pics sont des artéfacts de la numérisation. Une solution consiste à lisser la courbe avant de la dériver.

L'amplitude et la vitesse angulaire normalisé par ω sont représentées en fonction du temps sur la Figure 5a. Contrairement au signal *θ(t)*, la forme clairement triangulaire du signal *v(t)* atteste de la présence d'harmoniques impairs. Le graphe 5b représente la densité spectrale normalisée de *θ* et *v*. On rappelle que la densité spectrale est égale au carré de la norme de l'amplitude de la transformée de Fourier du signal. La transformée de Fourier est un outil mathématique très intéressant puisqu'il permet de visualiser les composantes sinusoïdales constitutive du signal. On trace plutôt la densité spectrale que l'amplitude spectrale afin de mieux faire ressortir les harmoniques par rapport au bruit. Comme l'atteste le modèle, on observe bien la présence d'harmoniques d'ordre 3 dont l'amplitude est respectivement de $(3\%)^2$ et $(10\%)^2$ du fondamental pour *θ* et *v*. Sur le spectre de la vitesse on observe même l'harmonique d'ordre 5. Cette harmonique trouve sa justification si l'on développe le modèle à l'ordre 3 en *ε*.

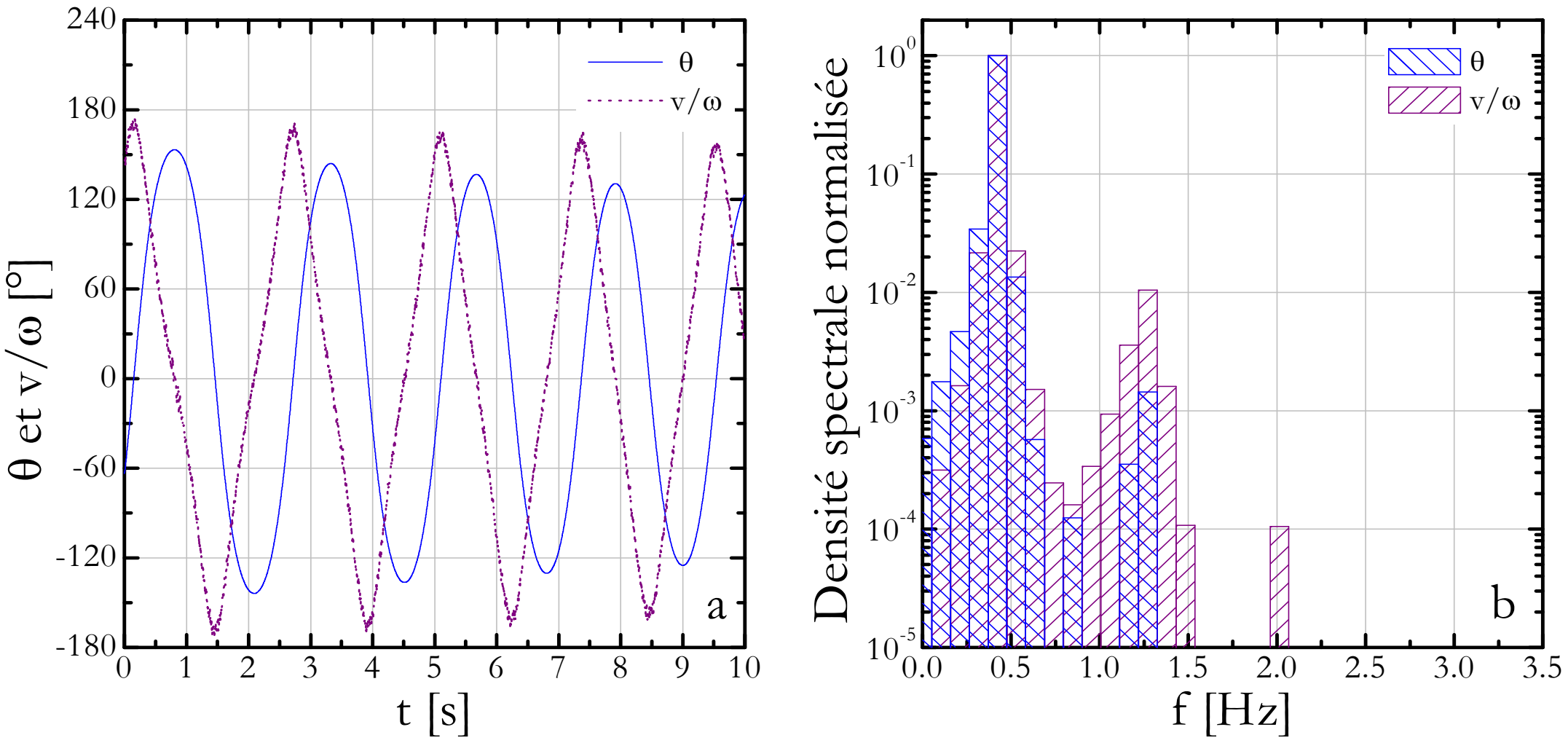

Figure 5 : (a) amplitude et vitesse angulaire divisé par ω en fonction du temps. La vitesse angulaire est obtenue en dérivant l'amplitude au préalable lissée sur 10 points. (b) densité spectrale normalisée par rapport au pic du fondamental de l'amplitude et de la vitesse angulaire. L'harmonique fondamental est situé à $f_0$=0.42Hz. Rq1 : Vu que l'on effectue une transformée de Fourier sur 10s l'axe des fréquences est gradué de 0.1Hz en 0.1Hz. Rq2 : la transformée de Fourier sous Latis Pro est améliorée par rapport à une transformée de Fourier normale en se sens qu'elle cale l'échantillonnage de l'axe des fréquence sur la fréquence du fondamental.

## 5. Conclusion

Nous avons mis en évidence les effets non linéaires du à l'anharmonicité du potentiel sur l'exemple du pendule. Nous avons montré que les effets non linéaires sont facilement observables sur la période pour peu que la période soit déterminée de façon précise. Le comportement de la période est bien interprété en développant l'énergie potentielle à l'ordre 4. La période est alors une fonction quadratique de l'amplitude des oscillations au moins dans une large gamme d'amplitude allant de 0° jusqu'à 30°. Ce développement prévoit aussi que des harmoniques d'ordre 3 apparaissent dans la solution de l'équation différentielle non linéaire. Les mesures des oscillations à des amplitudes angulaires élevées permettent de les mettre en évidence.

## Bibliographie

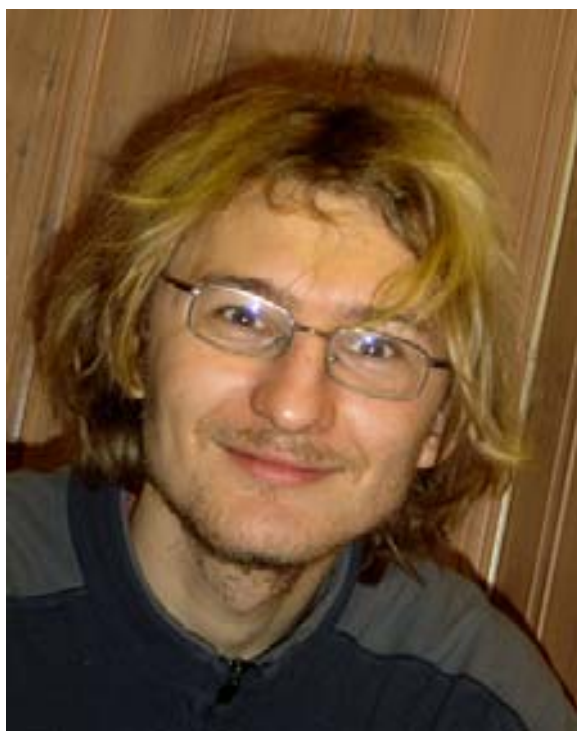

Thomas GIBAUD

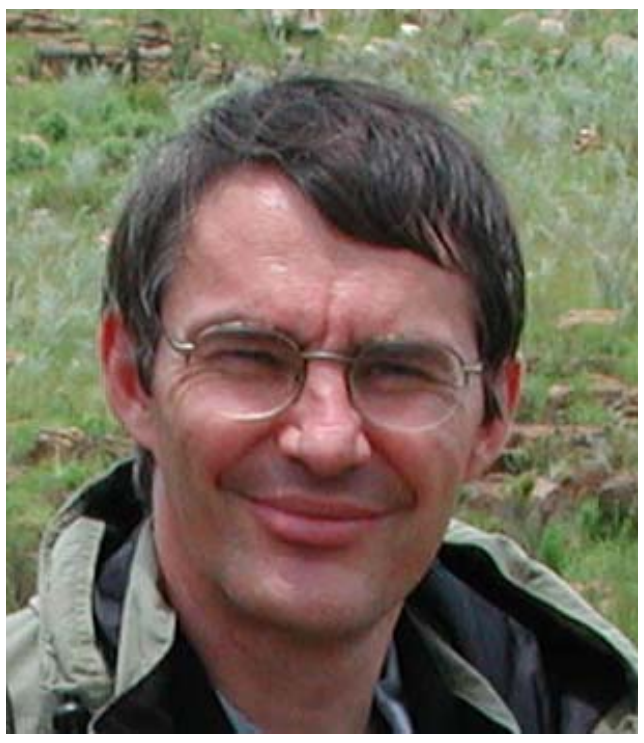

Alain GIBAUD